\documentclass[11pt,a4paper]{article}
\usepackage[utf8]{inputenc}
\usepackage{authblk}
\usepackage{indentfirst}
\usepackage{misccorr}
\usepackage{graphicx}
\usepackage{amsmath}
\usepackage{amssymb}
\usepackage{multicol}
\usepackage{bm}
\usepackage{longtable}
\usepackage{pdflscape}
\usepackage{epstopdf}
\usepackage{multirow}
\usepackage{adjustbox}
\usepackage{amsfonts}
\usepackage{fancyhdr}
\usepackage{ifthen}
\usepackage{fancyheadings}
\usepackage{setspace}
\usepackage[labelfont=bf]{caption}

\fancypagestyle{plain}{\chead{\scriptsize \texttt{Communications of BAO, Vol. 65 (2), 2018, pp. xxx-yyy}}}
\title{Blue straggler populations beyond the Milky Way}
\author{Richard de Grijs$^{1,2}$$^{\ast}$, Weijia Sun$^{3,4}$,
  Chengyuan Li,$^{1}$ Licai Deng$^{4}$}
\affil{\emph{\scriptsize$^{1}$Department of Physics and Astronomy,
    Macquarie University, Balaclava Road, Sydney, NSW 2109,
    Australia}\\
\emph{\scriptsize$^{\ast}$E-mail: richard.de-grijs@mq.edu.au}}

\affil{\emph{\scriptsize$^{2}$International Space Science
    Institute--Beijing, 1 Nanertiao, Zhonguancun, Hai Dian District,
    Beijing 100190, China}}

\affil{\emph{\scriptsize$^{3}$Kavli Institute for Astronomy and
    Astrophysics, Peking University, Yi He Yuan Lu 5, Hai Dian
    District, Beijing 100871, China}}

\affil{\emph{\scriptsize$^{4}$Key Laboratory for Optical Astronomy,
    National Astronomical Observatories, Chinese Academy of Sciences,
    20A Datun Road, Chaoyang District, Beijing 100012, China}}

\begin{document}
\pagestyle{empty}
\newpage
\pagestyle{fancy}
\label{firstpage}
\date{}
\maketitle

\begin{abstract}
Although the formation of blue straggler stars (BSSs) is routinely
attributed to stellar interactions in binary systems, the relative
importance of the direct collision and slow(er) stellar coalescence
formation channels is still poorly understood. We selected a sample of
24 Magellanic Cloud star clusters for which multi-passband {\sl Hubble
  Space Telescope} images are available to address this outstanding
question. We compiled a BSS database, containing both traditional and
evolved BSSs. We found a robust correlation between the number of BSSs
in a cluster's core and its core mass, $N_{\rm BSS, core} \propto
M_{\rm core}^{0.51 \pm 0.07}$, which supports the notion that BSS
formation is linked to a population's binary fraction. At low stellar
collision rates, the mass-normalised number of BSSs does not appear to
depend on the collision rate, which implies that the
coalescence-driven BSS formation channel dominates. Comparison with
simulations suggests that stellar collisions contribute less than 20\%
to the total number of BSSs formed.
\end{abstract}


\section{Blue straggler stars}

Blue straggler stars (BSSs) are among the most visible deviations from
the once generally accepted `simple stellar population' model of star
clusters. In essence, this model assumes that star clusters were
formed almost instantaneously from the same progenitor molecular
cloud, so that the nascent stars have almost identical ages and the
same chemical compostion. Translated into colour--magnitude space, the
Hertzprung--Russell diagrams of single-aged, single-metallicity star
clusters would exhibit narrow stellar evolutionary features, including
a tight main sequence and a well-defined, sharp main-sequence
turn-off, in turn leading to narrow subgiant and red giant
branches. BSSs, which are predominantly found at brighter magnitudes
and bluer colours than the main-sequence turn-offs of their host
clusters, do not fit well into this picture.

First discovered by Sandage (1953) in the outer regions of the old
Galactic globular cluster (GC) M3, many decades of research effort
has led to the realisation that BSS formation is intimately linked to
the evolution of stellar binary systems. In essence, the primary and
secondary components of the binary system evolve into a single object
with the combined mass of its progenitor components to gain a new
lease of life as an apparently rejuvenated BSS. The current consensus
is that BSSs can form in one of two ways, that is, through direct
collisions between two stars and through a slower process leading to
stellar coalescence. In the former scenario, when two low-mass stars
collide, they will form a gravitationally bound system that is subject
to rotation to satisfy the conservation of angular momentum. As both
stars merge, they release some of the collision products in the form
of debris, which eventually disperses. 

The resulting object is a more massive, rapidly rotating star that
appears blue because the stellar atmospheres of the progenitor stars
have been stirred up violently, allowing core materials to float up to
the surface layers. The merged collision product gradually heats up
under the effects of gravitational contraction and it eventually
expands to become a red giant-like star. Magnetic braking causes a
reduction in the star's rotation rate, allowing the star to shrink,
heat up again, and eventually settle as a slowly rotating BSS. In the
coalescence model, the resulting merger into a more massive star
occurs on a much longer timescale. The merger product is a rapidly
rotating BSS.

Observational evidence in apparent support of these two BSS formation
channels was first provided by Ferraro et al. (2009), who showed that
the colour--magnitude diagram of the core-collapse Galactic GC M30
exhibited two clearly separated `sequences' of stars in the area of
parameter space usually occupied by BSSs. The fainter, blue sequence
appeared to be an extension of the cluster's single-star main sequence
to brighter magnitudes (that is, including younger ages but offset by
$\Delta V \sim -0.4$ mag with respect to the younger extension of the
isochrone describing the GC's bulk stellar population), while the
bottom envelope of the brighter, red swarm of data points coincided
with the locus of the equal-mass binary sequence if extended to higher
masses than those defining the main-sequence turn-off. The natural
interpretation of these observations was that the blue BSS sequence
had resulted from stellar collisions triggered by the cluster's core
collapse, while the red main sequence and the stellar sample at
brighter magnitudes may have formed through stellar coalescence (but
see below). Note that this latter sample in Ferraro et al.'s (2009)
observations did not occupy a single, well-defined red sequence,
presumably because a fraction of the coalescence products had already
undergone further evolution, and thus they would have started to move
away from their birth environment in colour--magnitude
space. Alternatively, some of those objects could simply be unresolved
interacting binary systems involving non-main-sequence components that
have yet to merge.

Further evidence provided by Ferraro et al. (2009) suggested that the
collision products were mostly found in the GC's core envionment,
while the coalescence products occupied larger clustercentric
radii. Similar results were reported by Li et al. (2013) for the Large
Magellanic Cloud (LMC) GC Hodge 11, although their results were less
clear-cut. At the larger distance of the LMC, of order 50 kpc, the
effects of (back- or foreground) field contamination and stellar
crowding could significantly affect the quality and reliability of
photometric observations of dense star clusters in their outer and
core regions, respectively.

\section{Star clusters in the Magellanic Clouds}

Nevertheless, the Magellanic Clouds are better environments to study
the evolution of dense, massive star clusters than the Milk Way, for a
number of important reasons: (i) star clusters in the Magellanic
Clouds cover a large range in ages (by contrast, Milky Way GCs are
almost uniformly older than 10 Gyr); (ii) given their far southern
location in the sky, the Clouds are located far from the Galactic
plane, and hence they are negligibly affected by foreground extinction
(although a relatively small level of internal extinction still needs
to be dealt with); and (iii) the set of massive star clusters in the
Magellanic Clouds are less dense than their Galactic GC counterparts,
thus counteracting the distance effects (i.e., stellar crowding) to
some extent.

Despite these clear advantages of exploring the physical properties of
Magellanic Cloud clusters, their large distances still require us to
avail ourselves of high-resolution imaging observations with the {\sl
  Hubble Space Telescope} ({\sl HST}), particularly using the Advanced
Camera for Surveys (ACS)/ Wide-Field Channel (WFC) or the
Ultraviolet--Visible (UVIS) channel/Wide-Field Camera-3 (WFC3), to
resolve the bulk of their stellar populations. The resolution of $\sim
80$ milli-arcsec at optical wavelength attainable with these
instruments translates to $\sim 0.02$ pc at the distance of the
LMC. For our analysis we selected all intermediate-age (1--3 Gyr-old)
and old ($\sim$10 Gyr-old) clusters in the Magellanic Clouds for which
suitable {\sl HST} observations were available. Young massive clusters
were excluded, because the latter do not exhibit clear main-sequence
turn-offs, which thus complicates the selection of BSSs. Application
of these considerations resulted in a sample of 24 massive clusters
(for details, see Sun et al. 2018).

\begin{figure}[h]
\centering
\includegraphics[width=0.7\textwidth]{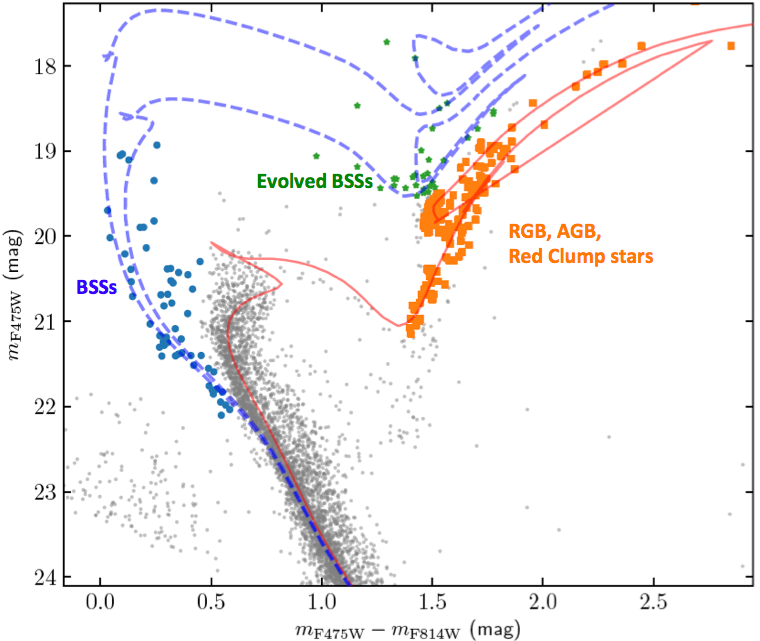}
\caption{Colour--magnitude diagram of the LMC star cluster NGC
  2173. BSSs, evolved BSSs, as well as red giant branch
  (RGB)/asymptotic giant branch (AGB)/red clump stars are marked with
  blue circles, green pentagons, and orange squares, respectively. The
  best-fitting PARSEC isochrones (Bressan et al. 2012) for the bulk
  stellar population (red solid line) and the BSS population (blue
  dashed line) are also shown. (Adapted from Sun et al. 2018.)}
\end{figure}

In a series of recent articles (Li et al. 2018; Sun et al. 2018), we
realised that while the best-fitting isochrones representing the bulk
stellar populations in our sample of Magellanic Cloud star clusters
also matched the ridgelines of their red giant branches very well,
many clusters exhibited significant numbers of stars blueward of the
blue envelopes defined by the red giant branch ridgelines and well
beyond the prevailing $3\sigma$ observational uncertainties. In fact,
isochrone fits to the cluster's BSS sequences appeared to adequately
match those bluer red giant branch-like stars, which led us to suggest
that the latter objects might be evolved BSSs: see Fig. 1. We
carefully validated this idea on the basis of Monte Carlo simulations
including the prevailing uncertainties, the number distributions, as
well as their spatial distributions (Li et al. 2018; Sun et
al. 2018). Henceforth, in relation to each of our sample clusters, we
will therefore refer to (i) our core BSS sample and (ii) our evolved
BSS sample. BSSs are found in all of our clusters, with BSS numbers in
their cores ranging from five to 70. Evolved BSSs can be resolved in
the intermediate-age clusters using optical filters, but they cannot
be disentangled easily from the red clump in old clusters.

Following Knigge et al. (2009) and Leigh et al. (2013), we explored
whether our Magellanic Cloud BSS numbers, $N_{\rm BSS}$, also
correlated with the core mass, $M_{\rm core}$, of their host
clusters. Knigge et al. (2009) and Leigh et al. (2013) reported a
relationship of the form
\begin{equation}
N_{\rm BSS, core} \propto M_{\rm core}^\alpha; \alpha \sim 0.5
\end{equation}
for samples of 57 and 30 additional Galactic GCs, respectively. Our 24
LMC clusters, on the other hand, yielded a significantly steeper
proportionality,
\begin{equation}
N_{\rm BSS, core} \propto M_{\rm core}^{\alpha_{\rm c}}; \alpha_{\rm
  c} = 0.66 \pm 0.07,
\end{equation}
and a Spearman coefficient of $\rho_{\rm S} = 0.84$ (Sun et
al. 2018). If we include the `evolved BSSs,' however, the power-law
index of this latter proportionality becomes $\alpha_{\rm c+e} = 0.51
\pm 0.07$ (and $\rho_{\rm S} = 0.80$), which implies a robust
correlation that is indeed very close to the previously published
Galactic GC results.

Knigge et al. (2009) suggested that the `sublinear' relation between
the number of BSS stars in a cluster and its core mass could be
explained as the result of binary interactions if the binary fraction,
$f_{\rm bin}$, also depends on the core mass (Milone et al. 2012),
\begin{equation}
f_{\rm bin} \propto M_{\rm core}^{-0.5}.
\end{equation}
However, Leigh et al. (2013) found that including the numbers of
binary stars in the cores of their sample of 30 additional Galactic
GCs did not strengthen the correlation, either because of the
empirical uncertainties affecting the derived binary fractions or
owing to the correlation's intrinsic dispersion; the resulting
degradation of the correlation further complicated efforts to
understand the BSSs' origins.

\begin{figure}[h]
\centering
\includegraphics[width=1\textwidth]{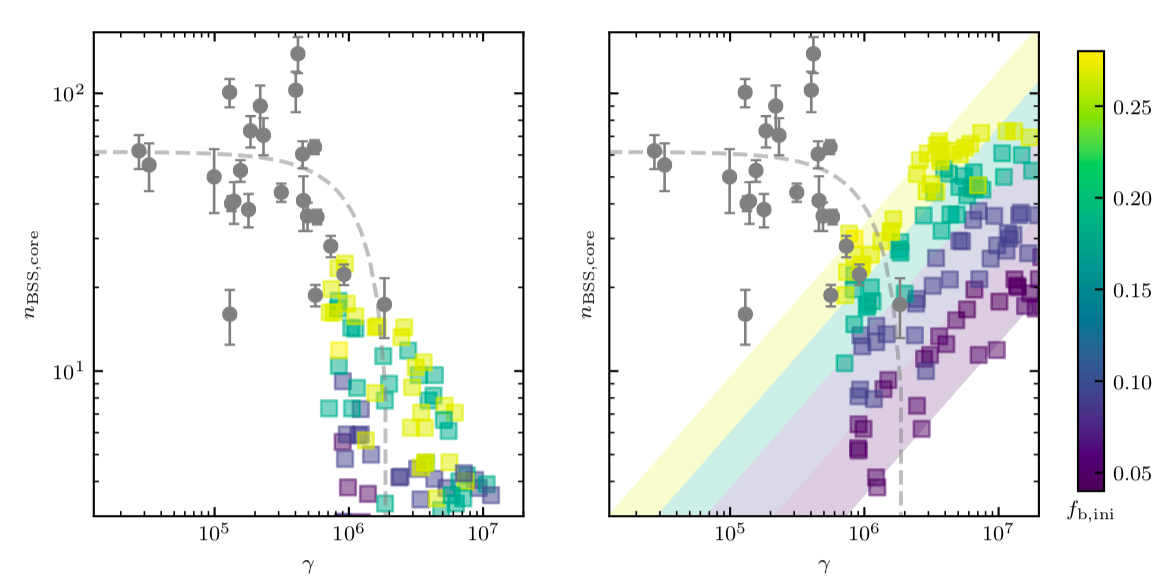}
\caption{Mass-normalised number of BSSs in the core ($n_{\rm
    BSS,core}$) as a function of the mass-normalised collision
  parameter, $\gamma$ (single star collisions only). Magellanic Cloud
  clusters are shown as the grey data points. The grey dashed line
  represents the best-fitting model of the form $n = C + a
  \gamma^b$. The simulation of Chatterjee et al. (2013) is
  overplotted. BSSs formed through the binary and collision channels
  in the simulation are shown in, respectively, the left- and
  right-hand panels as coloured squares, for different initial binary
  fractions ($f_{\rm b,ini}$). The shaded regions encompass the
  possible contributions from collisions in lower collision-rate
  environments. (Source: Sun et al. 2018.)}
\end{figure}

Therefore, we explored whether the numbers of BSSs, both in the
cluster cores and including the evolved counterparts, might be
correlated with the clusters' representative (annual) stellar
collision rates: see Fig. 2. We normalised both quantities by the
clusters' core masses to reduce any effects owing to a dependence of
the collision rates on the core masses of our clusters. It is clear
that in low collision-rate clusters, the collision rates do not depend
on core mass. At high collision rates, on the other hand, BSS
formation appears to be suppressed. We suspected that this could be
evidence of the importance of binary-mediated BSS formation.

Figure 2 also includes the results of the numerical simulations
performed by Chatterjee et al. (2013). The simulation results shown
in, respectively, the left- and right-hand panels highlight BSSs
formed through coalescence of binary systems and direct collisions,
for different initial binary fractions. The left-hand panel shows that
the mass-normalised number of BSSs generated from primordial binaries
in the Chatterjee et al. (2013) simulations declines toward higher
collision rates, as observed in our Magellanic Cloud cluster sample,
while the collision model predicts the opposite. This suggests that
binary disruption may be at work. Extrapolating, we estimate that the
share contributed by collisions in lower collision-rate environments
is less than 20\%, thus demonstrating a more significant contribution
from binary coalescence.

Although our results for 24 Magellanic Cloud clusters appear very
similar to those published previously for Galactic GCs, we emphasise
that the Magellanic Cloud clusters represent a special environment for
the study of BSS formation in star clusters. The underlying formation
channels might be different in the Magellanic Clouds compared with the
Milky Way. Indeed, the low stellar densities and low collision rates
in the Magellanic Clouds provides a great opportunity to resolve the
tension between observations and simulations in the interpretation of
the observed correlation between BSS number and core mass in Galactic
GCs.

\section*{\small Acknowledgements}
\scriptsize{RdG would like to thank the organisers of the conference
  `Instability Phenomena and Evolution of the Universe' (September
  2018), held in honour of Viktor Ambartsumian's 110th birthday, for
  their invitation to deliver this talk as well as for their partial
  financial support. This work was supported by the National Key
  Research and Development Program of China through grant
  2017YFA0402702 (RdG). We also acknowledge research support from the
  National Natural Science Foundation of China (grants U1631102,
  11373010, and 11633005). CL is supported by the Macquarie Research
  Fellowship Scheme.}

\end{document}